# A large-scale comparison of coverage and mentions captured by the two altmetric aggregators- Altmetric.com and PlumX

Mousumi Karmakar[a], Sumit Kumar Banshal[b], Vivek Kumar Singh[a,1]

[1]Department of Computer Science, Banaras Hindu University, Varanasi-221005, India
[2]Department of Computer Science, South Asian University, New Delhi-110021, India.

**Abstract:** The increased social media attention to scholarly articles has resulted in creation of platforms & services to track the social media transactions around them. Altmetric.com and PlumX are two such popular altmetric aggregators. Scholarly articles get mentions in different social platforms (such as Twitter, Blog, Facebook) and academic social networks (such as Mendeley, Academia and ResearchGate). The aggregators track activity and events in social media and academic social networks and provide the coverage and transaction data to researchers for various purposes. Some previous studies have compared different altmetric aggregators and found differences in the coverage and mentions captured by them. This paper attempts to revisit the question by doing a large-scale analysis of altmetric mentions captured by the two aggregators, for a set 1,785,149 publication records from Web of Science. Results obtained show that PlumX tracks more altmetric sources and captures altmetric events for a larger number of articles as compared to Altmetric.com. However, the coverage and average mentions of the two aggregators, for the same set of articles, vary across different platforms, with Altmetric.com recording higher mentions in Twitter and Blog, and PlumX recording higher mentions in Facebook and Mendeley. The article also analysed coverage and average mentions captured by the two aggregators across different document types, subjects and publishers.

## Introduction

The increased social media attention to scholarly articles has resulted in creation of platforms & services to track the social media transactions around them. These platforms track the various events about scholarly articles in social media (such as Twitter, Blog, Facebook), academic social networks (such as Mendeley, Academia and ResearchGate) and various other networks. These newer kinds of metrics (aka altmetrics) present useful insight about the importance and impact of the articles. Altmetrics are now being collected and analysed for a variety of purposes ranging from early impact assessment to measure of correlations between altmetrics and citations. Some studies have even tried to propose that altmetrics could be an alternative to citations for assessment of the impact of research (such as Haustein et al., 2014; Costas, Zahedi, & Wouters, 2015; Thelwall, 2017, 2018; Huang, Wang, & Wu, 2018; Thelwall & Nevil, 2018, Banshal, Singh & Muhuri, 2021). Due to importance and usefulness of the altmetric events, researchers are now increasingly using services of the altmetric aggregators for obtaining altmetric data for scholarly articles. Altmetric.com and PlumX are the two popular aggregators that track altmetric events around scholarly articles from a variety of platforms and provide the coverage and transaction data to researchers for various purposes. However, the outcome of any analysis of coverage and transaction data is likely to depend on the values provided by the aggregator used.

---

[1] Corresponding Author. Email: vivek@bhu.ac.in

Some recent studies have analysed the coverage and data quality of the two altmetric aggregators for different samples of scholarly articles. While several studies (such as Zahedi, Fenner, & Costas, 2015; Zahedi & Costas, 2018a, 2018b; Ortega, 2018a; Bar-Ilan, Halevi, & Milojević, 2019) focused on analyzing the agreement/ disagreement and differences in values of the altmetric scores across different altmetric data providers; some others (such as Meschede & Siebenlis, 2018; Ortega, 2018b, 2018c) attempted to measure the correlations among altmetric values from different aggregators. Two recent studies (Ortega 2019a, 2020b) tried to analyze the altmetric biases with respect to country, language and subjects; while another study (Ortega, 2019b) analysed the inconsistencies in data obtained from different aggregators. Most of the previous studies, however, either analysed small samples of scholarly articles. Only few recent studies (Ortega, 2019a, 2019b, 2020b) have used a large data sample of 100,000 Crossref DOIs for analysis.

This article, therefore, attempts to revisit the question of coverage variation, differences in altmetric scores and correlations in values provided by the two aggregators by analysing a very large sample of scholarly articles. We analysed a set of 1,785,149 publication records (17 times higher than the largest data sample used earlier), which constitutes the complete publication record for the world for the year 2016 indexed in Web of Science. The data for the year 2016 was taken for the reasons of stability of bibliometric records and also for availability of longer period (2016 to 2019) for the altmetric events to accrue. Given that a recent study (Karmakar, Banshal & Singh, 2021) has observed that altmetric events accrue for a longer time period, it was important to take altmetric data for a longer period for an accurate analysis of the differences in coverage and altmetric values of the two aggregators. Further, since the publication records comprised of the whole of Web of Science publication records for the world for a complete year, they represented different geographies, publication types, subjects, and publishers. Therefore, the study could also compare differences in coverage and altmetric values obtained from the two aggregators across different document types, subjects and publishers. Our study is thus unique in terms of use of a very large data sample, longer altmetric data period and variety of publication records analysed.

**Research Questions**

This article performed a comprehensive analysis of altmetric data for a very large set of scholarly articles and attempted to answer the following research questions:

*RQ1:* How much do the Altmetric.com and PlumX differ in their coverage of altmetric sources and scores?

*RQ2:* To what extent the altmetric mentions captured by Altmetric.com and PlumX correlate with each other?

*RQ3:* Whether the altmetric mentions captured by Altmetric.com and PlumX vary across different document types, subjects and publishers?

**A brief overview of the two altmetric aggregators**

An altmetric aggregator is typically a platform which tracks and accumulates various types of events from different social media, academic social networks and other platforms for scholarly articles. Altmetric.com and PlumX are the two-current popular altmetric aggregators. We present below a very brief overview of the two aggregators, emphasizing on their coverage of sources and the data collection/ update process.

**Altmetric.com**[2] is one of the first altmetric aggregator platforms, that originated in 2011 through the efforts of Euan Adie and support of Digital Science[3]. It crawls over different social media, news, and blog platforms to collect views, mentions, comments etc. around a scholarly article. Altmetric.com makes use of various scholarly identifiers like DOI, PubMed ID, ISBN, Handle, arXiv ID, ADS ID, SSRN ID, RePEC ID, URN, Clinical Trials.gov record etc. to track altmetric events around the scholarly articles. It employs different APIs and tools to track altmetric events from various sources[4]. Multiple sources of information for the same scholarly articles are cross-checked and summed up into a single entry in the respective platforms. For example, it uses GNIP API to track twitter mentions in real time[5]. A list of 9,258,010 unique tweeters is indexed, providing 125,793,313 tweets[6]. Twitter mentions include tweets, retweets, and quoted tweets but do not incorporate any tweets with links to news stories or blog posts. Similarly, Mendeley API is used to capture Mendeley reads, and the data is updated on a daily basis. Along with total reads, geographic locations, user's professional status and subject areas are also available[7]. In case of Facebook mentions, Altmetric.com captures only public pages, and the metric provided does not include likes, shares, and comments[8]. The data collection is done using Facebook Graph API. Currently, 297,808 Facebook sources are indexed in their database, with 4,635,255 total mentions[9]. The update is performed daily for public pages and weekly for prioritized popular pages[10]. In case of Blog mentions, the data is updated daily by RSS feed from a manually curated list of 9,653 blog sources[11] with direct links to scholarly identifiers[12]. Altmetric.com provides the facility to suggest any new blog source or Facebook page which is currently not indexed with them[12]. It also computes a weighted score based on mainly three factors- volume, sources, and authors - and provides a measure called 'Altmetric Attention Score'. This score is represented as a colorful donut on the article details page[13]. Altmetric.com offers various services for institutions, funding agencies, researchers, and agencies involved in research & development (R&D). It provides a free API with rate limit to researchers on request[14]. It also provides an environment to obtain data in different formats, like JSON, CSV etc.

---

[2] https://www.Altmetric.com/, accessed on 10 January, 2020
[3] https://www.digital-science.com/, accessed on 10 January, 2020
[4] https://help.altmetric.com/support/solutions/articles/6000240585-scholarly-identifiers-supported-by-altmetric, accessed on 25th February, 2021
[5] https://help.altmetric.com/support/solutions/articles/6000235926-twitter, accessed on 25th February, 2021
[6] https://www.altmetric.com/explorer/mention_sources?mention_sources_types%5B%5D=type%3Atw, accessed on 25th February, 2021
[7] https://help.altmetric.com/support/solutions/articles/6000236722-mendeley, accessed on 25th February, 2021
[8] https://help.altmetric.com/support/solutions/articles/6000235936-facebook, accessed on 25th February, 2021

**PlumX** is a suite of products launched by Plum analytics in 2016, initially with limited coverage[15]. Over time, these metrics evolved significantly in several ways and now it is a quite comprehensive altmetric aggregator. Plum Analytics considered as many as 67 different types of outputs to be tracked, which are named as 'artifacts'. These artifacts include scholarly articles, books, book chapters, conference articles. In addition, it also includes speeches, visual arts, images, figures etc. Currently, PlumX track metrics over 180 million research outputs. It covers a very wide variety of social platforms, such as Twitter, Facebook, YouTube; online knowledge sharing mediums, such as StackExchange, Wikipedia, Github; and bibliographical data-based sites, such as Scopus, SciELO, RePEcetc. It organizes the captured data in five different types of metrics- usages, captures, mentions, citations, and social media[16]. The data for artifacts is collected from multiple platforms using different methods and tools. These include data provider APIs, third party APIs, FTP data transfers, OAI-PMH harvesting, Web crawlers, and RSS feeds[17]. For example, it updates twitter metrics in real time using GNIP API, whereas Facebook Graph API is used with daily update for Facebook data (Zahedi & Costas, 2018b). The Facebook metric provided by PlumX includes shares, likes, and comments along with the interactions in user's closed network[18]. It is not very clear how PlumX collects information about Mendeley readers and Blog mentions. However, since Mendeley and PlumX both are owned by Elsevier, one may expect that PlumX may be getting direct access to Mendeley data (Zahedi & Costas 2018b). Various blog sources are mined to collect the blog mention data, but the list of blog sources is not disclosed in public domain. The data harvesting is updated on different time periods, ranging from daily to monthly basis, based on the different

---

[9] https://www.altmetric.com/explorer/mention_sources?mention_sources_types%5B%5D=type%3Afb, accessed on 25th February, 2021
[10] https://help.altmetric.com/support/solutions/articles/6000240275-attention-sources-update-frequency-and-collection-methods, accessed on 25th February, 2021
[11] https://www.altmetric.com/explorer/mention_sources?mention_sources_types%5B%5D=type%3Ablog, accessed on 22nd February, 2021
[12] https://help.altmetric.com/support/solutions/articles/6000235927-blogs , accessed on 25th February, 2021
[13] https://www.Altmetric.com/about-our-data/the-donut-and-score/, accessed on 10 January, 2020
[14] https://www.Altmetric.com/research-access/, accessed on 10 January, 2020
[15] https://plumanalytics.com/learn/resources/plum-analytics-metrics-audit-log/, accessed on 10 December, 2019
[16] https://plumanalytics.com/learn/about-metrics/, accessed on 10 December, 2019
[17] https://plumanalytics.com/niso-altmetrics-working-group-on-data-quality/, accessed on 10 December, 2019
[18] https://plumanalytics.com/plumx-facebook-altmetrics-measure-up/, accessed on 25th February, 2021

licensing policies of the harvested platforms. Plum Analytics refreshes PlumX in every 3-4 hours to keep it most updated. The data can be accessed through end-user interfaces, widgets, and APIs of Plum analytics.

The two altmetric aggregators, Altmetric.com and PlumX, provide metrics based on data collected from various social media, bibliographic and policy document sources. **Table 1** lists a total of 33 social media sources tracked by the two aggregators. Out of these 33 sources, 14 sources are captured in Altmetric.com, whereas PlumX tracks 28 sources. The platforms/ sources tracked by Altmetric.com are Twitter, Facebook, Youtube, Reddit, F1000, Blog, Mendeley, Stack Overflow, Wikipedia, News, CiteULike, LinkedIn, Google+, and Pinterest. Out of these 14 sources, PlumX tracks most except five sources- F1000, LinkedIn, Google+, Stack Overflow, and Pinterest. PlumX additionally tracks 19 social media sources that are not tracked by Altmetric.com. These sources are bit.ly, Figshare, Github, Slideshare, SoundCloud, SourceForge, Vimeo, Stack Exchange, Goodreads, Amazon, Delicious, Dryad, Dspace, SSRN, EBSCO, ePrints, AritiiRead eBooks, Ariti Library, WorldCat. **Table 2** lists the bibliographic and policy document sources tracked by the two aggregators. Here, PlumX has a better coverage and also provides citation metrics. PlumX covers a total of 16 sources whereas Altmetric.com tracks only 6 sources. Thus, tables 1 and 2 suggest that PlumX covers more varied sources as compared to Altmetric.com. The two aggregators also differ in their strategies to capture altmetric events and update frequency. The study by Zahedi & Costas (2018b) provides a good overview of the main methods of collecting, tracking, and updating metrics by some popular altmetric aggregators.

**Related Work**

Several previous studies tried to analyze the data from different altmetric aggregators for different purposes ranging from assessing their accuracy to finding how much they agree on altmetric counts for same set of scholarly articles.

Zahedi, Fenner & Costas (2015) explored the agreement/disagreement among metric scores across three altmetric providers namely, Mendeley, Lagotto, and Altmetric.com. They analysed 30,000 DOIs for the year 2013 in five common sources and analysed possible reasons for the differences. They found that Altmetric.com reports more tweets as compared to Lagotto and concluded that the data capture procedure of Altmetric.com, which includes tweets, public retweets, and comments in real-time, could be a probable reason for such differences. In later studies, Zahedi & Costas (2018a, 2018b) have analysed 31,437 PloSOne DOIs and explored the differences in metrics provided by four aggregators Crossref Event Data (CED), Altmetric.com, Lagotto, and Plum Analytics. They focused on the process of data collection used by different aggregators and that how different aggregators define metrics from the data collected. The results showed that Mendeley ($r>0.8$) and Twitter ($0.5 \leq r \leq 0.9$) have good agreement across aggregators, whereas Facebook ($0.1 \leq r \leq 0.3$) and Wikipedia ($0.2 \leq r \leq 0.8$) have the lowest agreement. They attributed this to the methods of tracking and processing data. For example, the effect of direct data collection or collection through third-party APIs, aggregation of data based on different versions, identifiers, types, etc., and the impact of frequency of update, etc. They recommended that one should not rely only on the aggregators showing a higher count for the metric.

Meschede & Siebenlist (2018) explored about the relationship between the metrics across (inter-correlation) two aggregators PlumX and Altmetric.com, as well as between the metrics within the aggregator itself (intra-correlation). They analysed sample of 5,000 journal articles from six disciplines ('Computer Science, Engineering and Mathematics', 'Natural Sciences', 'Multidisciplinary', 'Medicine and Health Sciences', 'Arts, Humanities and Social Sciences' and 'Life Sciences') and analysed them for the eight common sources ('Facebook', 'Blogs', 'Google+', 'News', 'Reddit', 'Twitter', 'Wikipedia' and 'Mendeley') in both aggregators. The study showed that PlumX has higher overall coverage (99%) of the data chosen for analysis as compared to Altmetric.com (39%). The intra-correlation between the metrics within the same platforms are weak. They further observed that PlumX and Altmetric.com are highly inter-correlated in terms of Mendeley and Wikipedia (with correlation coefficient values 0.97 and 0.82 respectively) but weakly correlated for other sources- Facebook (0.29), Blogs (0.46), Google+ (0.28), News (0.11), Twitter (0.49), and Reddit (0.41).

Ortega (2018a) analysed the difference in altmetric indicator counts in Crossref event data, Altmetric.com, and PlumX, using a sample 67,000 papers. For each platform, the difference in metrics across aggregators was quantified in terms of counting differences. Counting difference was computed by taking the sum of the differences in metrics provided by two aggregators at the document level and dividing by the number of publications that have non zero altmetric events and occur in both aggregators. They concluded that different aggregators should be used for data from different platforms, such as PlumX for Mendeley reads and Altmetric.com for tweets, news & blogs. In another study, Ortega (2018b) has grouped different altmetrics into three groups: social media, usage, citations using principal component analysis (PCA). In this study data from Altmetric.com, Scopus and PlumX for a set of 3,793 articles published in 2013, was used. Considering that the earlier studies provided evidence that some specific aggregators perform better for some specific data sources; they collected different indicators from different aggregators. These included tweets, Facebook mentions, news, blogs etc. from Altmetric.com; citations from Scopus; Wikipedia mentions from CED; and views & Mendeley reads from PlumX. Results showed that instead of using a single metric, such as altmetric score, one should consider the relatedness of metrics and their impact across different disciplines for evaluating research. In another study, Ortega (2018c) examined the emergence and evolution of five altmetrics (download, views, tweets, readers, and blog mention) along with bibliometric citations from the publication date of a document. The study also investigated the evolution of the relationships among these metrics by analyzing 5,185 papers from PlumX on a month-to-month basis. The results showed that in a document's entire life cycle, altmetric mentions are fast appearing ones, whereas citations appearance is slow. Based on the relationship analysis of metrics, the study suggested that the reader counts influence citations.

Bar-Ilan, Halevi, & Milojević (2019) have analysed altmetric data of 2,728 JASIST articles and reviews, provided by Mendeley, Altmetric.com and PlumX in two different points of time 2017 and 2018. They observed increase in overlap in coverage of documents with Mendeley reader across the three sources over time. There were 874 papers commonly covered in all sources in 2017, which increased to 1,021 papers in 2018. Further an increase in Mendeley reader counts and citations was also observed. They suggested using more than one aggregator to obtain altmetric indicators and to compare them in order to get reliable altmetrics.

A series of recent studies (Ortega, 2019a, 2019b, 2020b) analysed coverage of news and blog sources in three aggregators namely, Crossref event data, Altmetric.com, and PlumX, by taking 100,000 Crossref DOIs. The results showed that, the overlap of these sources across aggregators are comparatively low in numbers (Ortega, 2019a). As for example, Altmetric.com has a higher coverage for blog (37.8%) but only 7.8% of the publication set is commonly covered in the three aggregators. The coverage in one aggregator might be high for the same set of articles but the lower overlapping ratio shows that the sources covered in aggregators vary widely. Ortega (2020b) explored altmetric biases with respect to country, language and subjects with a dataset of 100,000 DOIs. Author has retrieved the sources which covered the randomly selected publication set and categorized them based on their regions, language and interest level. It was shown that, Altmetric.com is the most heterogeneous aggregator geographically and linguistically. However, PlumX had more coverage towards local news events, particularly for USA. Their conclusion served as evidence that English is the most prevailing language. From the same dataset, Ortega (2019b) extracted the blog and news links to verify the validity, coverage, and presence of the tracked blog mentions and news mentions of the scholarly articles. There were 51,000 news & blog links found in this extraction process, which were explored for their existence and it was found that almost one-third of the links are broken. This elaborate longitudinal study concluded that these mentions should be audited periodically as the aggregators are dependent on third-party providers.

Ortega (2020a) has performed a meta-analysis over a set of 107 altmetric articles related to five altmetric aggregators, namely, Altmetric.com, Mendeley, PlumX, Lagotto, and ImpactStory, published during 2012-2019. The dataset consisted of papers that had either computed or published data useful in the computation of three metrics: coverage, platform wise coverage, and average mentions. The usage percentage of all the aggregators was explored. Almetric.com (54%) was found to be the most prevalent provider, followed by Mendeley (18%) and PlumX (17%). The analysis showed that Altmetric.com tracks more events for Twitter, News, and Blogs whereas PlumX performs well in Facebook and Mendeley platforms. The results exhibited gradual increase in tweet capture by PlumX.

**Data**

Since the focus of the work is comparing the coverage and scores of altmetrics provided by the two popular aggregators- Altmetric.com and PlumX, we analysed the variation in coverage and scores for a very large set of articles. The whole of the world's research output for the year 2016 as indexed in Web of Science (WoS) was downloaded for the analysis. The download was performed in the month of Sep. 2019. The data for the year 2016 was taken for the reasons of stability of bibliometric records and also for availability of longer period (2016 to 2019) for the altmetric events to accrue. Since, WoS does not allow downloading data above 100,000 records, therefore the data is collected based on Web of Science Categories (WC). The WC based data collection has an inherent problem of duplicity since in WoS a paper is generally tagged under many WCs. Due to this duplicity, the downloaded data comprised of 3,545,720 publication records. This data comprised of the standard 67 fields- including TI (title), PY (publication year), DI (DOI), DT (document type), SO (publication name), DE (author keywords), AB (abstract) etc. This data was then processed to remove duplicate and incomplete records. After this process, we were left with 1,785,149 publication records with DOI.

The next step was to obtain altmetric data for these publication records from the two aggregators- Altmetric.com and PlumX. In order to obtain altmetric data from Altmetric.com, a DOI look up was performed for all the DOIs in the WoS data. Out of the 1,785,149 publication records, a total of 902,990 records are found to be covered by Altmetric.com, which is about 50.58% of the total data. Altmetric.com had 46 fields in the data, including DOI, Title, Twitter mentions, Facebook mentions, News mentions, Altmetric Attention Score, OA Status, Subjects (FoR), Publication Date, URI, etc. Out of this, we mainly used data for Twitter, Facebook, Mendeley, and Blog platforms. The data from Altmetric.com was downloaded in the month of Oct. 2019. For obtaining data from PlumX, we contacted PlumX team to provide us with access to PlumX data for the 1,785,149 publication records, as we did not have access to PlumX. The PlumX team created a dashboard access for us for the concerned publication records. Out of the 1,785,149 publication records, a total of 1,661,477 publication records were found covered in PlumX, which constitutes about 93.07% of the total data. PlumX provides metrics in five categories from a wide range of source platforms. The PlumX data was downloaded in the month of Nov. 2019. This data included fields like DOI, Title, Year, Repo URL, Researcher Name(s), Captures:Readers:Mendeley, Social Media:Tweets:Twitter, Social Media:Shares, Likes & Comments:Facebook etc. This data also has a field named Plum stable url which redirects to the page from where one can get the actual tweets, blogs etc. We have focused our attention mainly to data for four platforms- Twitter, Facebook, Mendeley, and Blog platforms- from both the aggregators.

**Methodology**

The altmetric data obtained from the two aggregators for WoS publication records was analysed on six aspects: variation in coverage, difference in magnitude of mentions, correlations in mention values, variation across document types, variation by subjects and variation across publishers.

*First of all*, the coverage in different platforms of scholarly articles by the two aggregators was compared. The altmetric data for the articles in consideration was obtained from the two aggregators corresponding to the four platforms: Twitter, Facebook, Blogs and Mendeley. The percentage of articles covered in the four platforms as per the data from the two aggregators was identified and difference in coverage was compared.

*Secondly*, the magnitude of mentions in the four platforms for the articles was analysed and the difference in magnitude of mentions in the data drawn from the two aggregators was computed. Statistical measures (mean and median) were computed for the differences in values from each of these platforms.

*Thirdly*, the correlation between mention values from different platforms, as drawn from the two aggregators, was computed. For computing correlations, the options were to compute Pearson Correlation or Spearman Rank Correlation. However, as it has been observed in previous studies (such as Thelwall & Nevill, 2018) that the altmetric data are highly skewed, therefore, we have used Spearman Rank Correlation, which is more suitable for such skewed data. The Spearman Rank Correlation Coefficient (SRCC) was computed between the different types of mentions available from the two aggregators. The built-in function '*corr*' available in pandas module of python programming language was used for this purpose, value 'spearman' passed as parameter to the function. The value of SRCC lies between -1 to +1, with positive values indicating positive correlation, value of 0 indicating no correlation, and negative value indicates negative correlation.

*Fourthly*, the difference in coverage and mentions captured by the two aggregators for the different platforms was analysed across different document types. The document type of articles was taken from the 'DT' tag in the WoS record file. The values for 'DT' include journal articles, proceedings paper, book chapters, reviews, book reviews, editorial material etc. The variation in coverage levels and magnitude of mentions for different platforms was thus obtained for these document types.

*Fifthly*, the difference in coverage and mentions across different subjects was computed by grouping the publication records into different subjects. Each publication record was grouped into one of the fourteen major subject categories as per the scheme proposed in (Rupika et al., 2016). The Web of Science Category (WC) field information for each publication record is seen and based on its value the publication record is assigned to one of the fourteen broad subject categories. These fourteen broad subject categories are as follows: Agriculture (AGR), Art & Humanities (AH), Biology (BIO), Chemistry (CHE), Engineering (ENG), Environment Science (ENV), Geology (GEO), Information Sciences (INF), Material Science (MAR), Mathematics (MAT), Medical Science (MED), Multidisciplinary (MUL), Physics (PHY) and Social Science (SS). This subject grouping into 14 broad categories instead of more than 255 categories of Web of Science, made the analysis more manageable and easier to understand. The variations in coverage and magnitude of mentions were thus computed for publication records in each of the fourteen subject groups.

*Finally*, difference in journal coverage and mentions was computed for 16 most frequent and well-known publishers namely 'Springer', 'Nature', 'PLoS', 'Elsevier', 'IEEE', 'Wiley', 'Taylor &Francis', 'ACM', 'IOP', 'Oxford University Press', 'Sage', 'Hindawi', 'Cell Press', 'MDPI', 'Cambridge University Press', and 'Emerald' across two aggregators. The publisher information was obtained from the 'PU' tag of WoS records. In the PU field same publishers are present in different forms. For e.g. Elsevier have variants such as Elsevier Science Ltd, Elsevier Masson, and Elsevier Science Bv. All such variants represent same parent publisher Elsevier. Therefore, all such variants of any publisher were combined through partial string match strategy verified later manually. This way all publication records for different publishers are obtained and differences in their coverage and in mention counts are computed across the two aggregators.

Results

The altmetric data captured by the two aggregators for the set of WoS articles was analysed to identify differences in coverage and magnitude of mentions across different platforms. The differences in mentions captured by the two aggregators were also analysed across different subjects, document types and publishers. The subsections below present these results.

*Difference in coverage of the two aggregators*

First of all, the difference in coverage of the two altmetric aggregators was identified. It was observed that out of the total set of 1,785,149 publication records, a total of 902,990 records are found to be covered by Altmetric.com (which is about 50.58% of the total data), and a total of 1,661,477 publication records were found covered in PlumX (which constitutes about 93.07% of the whole data). **Figure 1** shows the overlap of coverage of the two aggregators. It can be seen that a total of 879,981 articles are commonly covered by the two aggregators. About 97.5% of articles covered by Altmetric.com are also covered by PlumX, whereas Altmetric.com covers only 53% of articles tracked by PlumX. The PlumX aggregator has 47% of articles uniquely covered. Thus, it is observed that PlumX has a higher overall coverage of articles (including uniquely covered articles) as compared to Altmetric.com, indicating wider coverage of PlumX.

We have tried to find out whether the difference in coverage of the two aggregators is similar across different platforms. **Figure 2** shows a bar chart of article coverage of the two aggregators for four different platforms- Twitter, Facebook, Mendeley and Blog. It can be observed that PlumX has a better coverage for Mendeley platform, whereas Altmetric.com has an edge over PlumX in coverage in the Twitter and Blog platforms. The coverage for Facebook platform of the two aggregators is almost similar. The magnitude of coverage difference between the two aggregators is highest for Mendeley and lowest for Facebook. Thus, while PlumX has an overall higher coverage of articles, Altmetric.com has better coverage in two of the four platforms analysed.

*Difference in magnitude of mentions*

The mean and median values of number of mentions for the four platforms as tracked by the two aggregators was computed. **Table 3** shows the number of articles tracked by the two aggregators across the different platforms, along with the mean and median values of mentions. It can be observed that the mean value of mentions for Twitter and Blog platforms is higher in Altmetric.com, whereas the mean value of mentions for Facebook and Mendeley is higher in PlumX. In case of Facebook platform, PlumX platform has significantly higher value of mean and median of mentions as compared to Altmetric.com. It may, however, be noted that the mean/ median values are of different number of articles tracked by the two aggregators.

A more useful comparison of the value of mentions would require comparing the mentions for the commonly covered set of articles by the two aggregators. Therefore, we have compared the values of mentions captured by the two aggregators across different platforms for the same set of commonly covered articles. The difference in mention value for the papers in Altmetric.com and PlumX is computed. **Figure 3** shows the mean of differences in mentions in the four platforms as tracked by the two aggregators. It is observed that in case of Twitter and Blog platforms, the mean value of differences is positive indicating that Altmetric.com captures higher number of mentions as compared to PlumX. In the Facebook and Mendeley platforms, PlumX is found to have higher number of mentions tracked as compared to Altmetric.com.

In order to gain further insight into the difference in magnitude of mentions, we have also plotted the frequency of differences in mentions across different platforms by the two aggregators. **Figure 4 (a) – (d)** present the frequency values of differences in mentions in the two aggregators for the Twitter, Facebook, Mendeley and Blog platforms, respectively. **Figure 4(a)** shows the histogram for the differences in Twitter platform. These differences are for 565,445 commonly covered articles for Twitter platform, with at least one tweet captured in both the aggregators. It can be seen that a good percentage (approximately 50%) of papers has tweet difference equal to zero, indicating that both platforms record same number of tweets for these papers. However, the slope is inclined towards the positive side, indicating that Altmetric.com captures more tweets for a good number of the articles as compared to PlumX. We looked at some examples to verify this and found this valid. One example paper titled "When the Great Power Gets a Vote: The Effects of Great Power Electoral Interventions on Election Results" has 24,318 tweets captured by Altmetric.com, whereas PlumX captured only 493 tweets for this paper.

**Figure 4(b)** shows the histogram of article level differences in Facebook platform. Here the plot is created for 71,437 commonly covered articles in Facebook platform, that have non-zero Facebook mentions in both aggregators. It is observed that in approximately 16% of the articles, the difference in mentions is zero. However, the slope is clearly inclined towards the negative side, indicating that PlumX captures more mentions per article as compared to Altmetric.com for majority of the articles. One example article to mention would be the article titled "The terrorist inside my husband's brain.", which has 41,214 mentions captured by PlumX but only 105 mentions captured by Altmetric.com.

The histogram for the differences in mentions in the Mendeley platform is shown in **Figure 4(c).** Here, the plot is made for a total of 830,520 commonly covered articles in Mendeley platforms, that have at least one read recorded in both the aggregators. In this case too, it is seen that pattern is inclined more towards the negative side, indicating that PlumX captures more reads per article than Altmetric.com for majority of the articles. About 25% of articles have the same number of mentions recorded by the two aggregators. One example article would be article titled "Mastering the game of Go with deep neural networks and tree search" that has 39,621 records captured by PlumX but only 7,900 reads captured by Altmetric.com.

**Figure 4(d)** plots the histogram of the differences in Blog mentions for the 14,387 commonly covered articles in Blog platform, with non-zero mentions captured by both the aggregators. In this case, it is observed that more than 40 % articles have this difference equal to zero. The pattern, however, is inclined towards the positive side, indicating that Altmetric.com captures more mentions as compared to PlumX for a good number of articles. One example article to mention would be an article titled "Planet Hunters IX. KICÂ 8462852 Â– where's the flux?" that has 95 mentions captured by Altmetric.com but only 12 mentions captured by PlumX.

Thus, a perusal of the figures 4 (a) to (d) indicate that Altmetric.com captures more mentions per article in case of Twitter and Blog platforms, whereas PlumX captures more mentions per article for the Mendeley and Facebook platforms.

*Correlations in mentions*

We have also computed correlation between the mention values for different platforms across the two aggregators. The *Spearman Rank Correlation Coefficient (SRCC)* between mentions is computed for the articles commonly covered by the two aggregators. **Table 4** shows the SRCC values in the article-mentions across the two aggregators. It can be observed that the correlation values for Twitter and Mendeley platforms are 0.823 and 0.95, respectively, indicating strong correlation. In case of Facebook and Blog platforms, these values are 0.272 and 0.424, respectively, indicating weaker correlation. Thus, it can be said that there is more agreement in mention-based ranks of articles in Twitter and Mendeley platforms, between the two aggregators. The mention values differ more in the other two platforms (Facebook & Blog). The inter-platform correlations across the two aggregators are also shown, all of which are less than 0.5, indicating weak positive rank correlations across the platforms in the two aggregators.

*Variations across document types*

It would be interesting to check whether the coverage and mentions in different platforms as captured by the two aggregators vary across different document types. We have, therefore, analysed the coverage and mentions for the articles of different document types. These document types correspond to Article, Book, Book Chapter, proceedings paper and Review, as defined by the Web of Science. **Table 5** shows the coverage and average mentions for Twitter, Facebook, Mendeley and Blog platforms for the different document types. It is observed that Altmetric.com has better coverage in Twitter and Blog platforms for almost all document types. In case of average mention values too, Altmetric.com has higher values across almost all document types in Twitter and Blog platforms. The PlumX platform is found to have better coverage and average mention values across almost all document types in the Facebook and Mendeley platforms. Thus, looking at the results across document types, it is seen that the overall trend of better coverage of Altmetric.com for Twitter and Blog and of PlumX for Facebook and Mendeley holds valid across different document types.

*Variations across subjects*

The variations in coverage and mentions between the two aggregators is also analysed across different subjects. We have used the data grouping into fourteen broad subjects. **Table 6** presents the coverage and average mention values in the four platforms. It is observed that in the Twitter platform, Altmetric.com has better coverage and higher mention values than PlumX for almost all subjects. In case of Facebook platform, PlumX has higher mention values for almost all subjects. The coverage, however, is not higher for PlumX in Facebook for all subjects as subjects like MED, AH, SS, BIO, and AGR has higher coverage by Altmetric.com. In Mendeley platform, PlumX has higher coverage in all subjects, but in terms of reads, PlumX captures higher reads only for MED, SS, BIO, GEO, and MUL subjects. In case of Blog platform, Altmetric.com has better coverage than PlumX across almost all subjects and the average mention values of Altmetric.com are higher except for PHY, ENV, MAT and ENG subjects. Thus, the analysis of data across different subjects shows an overall trend of better coverage of Altmetric.com of Twitter and Blog and PlumX of Facebook and Mendeley, except in case of some subjects where slightly different patterns are observed.

*Variations across Publishers*

We have also tried to see if the patterns of variations in coverage and mentions in the two aggregators change across different publishers. In order to analyse this, articles for the 16 most frequent publishers in the publication data are identified and analysed. **Table 7** present the coverage and average mention values for data for these publishers in the four platforms for the two aggregators. In terms of number of journals for which data is covered, the PlumX aggregator has an edge over Altmetric.com. For example, PlumX covers 76 more journals of Springer than Altmetric.com, 34 more journals of Elsevier and 31 more journals of Taylor & Francis. It can be further observed that PlumX covers more than 90% of publication records for all Publishers except for Cambridge Univ Press (84.1%), whereas coverage of Altmetric.com varies between 25% to 86%. Altmetric.com has minimum coverage of about 25% for IEEE and highest coverage of 86.88% for PLoS publications. In terms of coverage and mentions for the four platforms, it is found that Altmetric.com has higher coverage in Twitter for almost all publishers. In Facebook platform, Altmetric.com shows higher coverage for all publishers except PLoS, Hindawi, and MDPI. In Mendeley platform, the coverage and average reads captured by PlumX are higher for all publishers except Springer, IEEE, Taylor & Francis, ACM and Hindawi. In case of Blog platform, Altmetric.com has in general better coverage and average mention values than PlumX, though for IEEE, IOP, Hindawi and Emerald publishers, the PlumX aggregators capture more mentions. Thus, in general it is observed that Altmetric.com has better coverage of Twitter and PlumX has better coverage of Facebook, irrespective of the publisher. However, in case of Mendeley and Blog platforms, the coverage and mention values of the two aggregators vary for different publishers, with no definite pattern observed across all publishers.

**Discussion**

The article tried to present a comparative analysis of coverage and mentions captured by the two well-known altmetric aggregators, namely, Altmetric.com and PlumX, for four platforms-Twitter, Facebook, Mendeley and Blog. The variations in coverage and mention values captured by the two aggregators for different platforms is analysed in general and also across different document types, subjects and publishers.

The results show that PlumX has an overall higher and wider coverage than Altmetric.com, with PlumX tracking about 93% articles as compared to Altmetric.com tracking about 50% articles. This is also confirmed by the findings of some previous studies (Meschede & Siebenlist, 2018; Ortega, 2018b; 2019a; Zahedi & Costas, 2018b), which in general found that PlumX has higher coverage of articles, with close to 95% articles of the data set being tracked.

The coverage difference, however, is not same across all the four platforms analysed. For example, in the case of Twitter platform, Altmetric.com is found to have better coverage and more average tweets than PlumX. This is interesting to observe given that both the aggregators collect information using the same API (GNIP) and that both update the data in real time. The previous studies by Ortega (2018a) have also found a similar pattern for Twitter, with Altmetric.com performing better than PlumX. One probable reason for such difference might be the fact that the two aggregators track articles by using different scholarly identifiers. Zahedi & Costas (2018b) have mentioned that the use of different scholarly identifiers could be one of the reasons for such variations. They have also reported that different handling policies of deleted tweets, protected tweets and suspended accounts by the two aggregators could be another reason for the differences observed. For example, PlumX removes all such tweets whereas Altmetric.com keep on counting the deleted ones. Thus, different handling of the tweet data appears to be the main reason for variations in coverage and values of tweets captured by the two aggregators.

In case of Mendeley platform, PlumX is found to have better coverage and higher number of reads as compared to Altmetric.com. Zahedi & Costas (2018b) have also observed similar pattern and reported that PlumX aggregates the reader counts of different versions of the same research output, resulting in overall higher Mendeley reads. Another possible reason may be the fact that PlumX and Mendeley are from the same parent company and hence they may have been better integrated together, which may allow more accurate and real time update of the data. Thus, a possibly better integration and recording counts of different versions of the same article by PlumX could be the main reasons for higher coverage of Mendeley by PlumX.

In case of Facebook platform, the coverage levels of the two aggregators are found to be quite similar, though PlumX has an edge over Altmetric.com. Ortega (2020a) has also found that in collecting mentions from Facebook and Mendeley, Altmetric.com has performed poorly as compared to PlumX. One possible reason for Altmetric.com recording slightly lesser Facebook mentions is that Altmetric.com captures only public pages[19] and excludes likes, shares, and comments, whereas PlumX includes shares, likes, and comments along with the interactions in user's closed network. A similar argument is also given by Zahedi & Costas (2018b) behind the low Facebook mentions captured by Altmetric.com.

---

[19]https://help.altmetric.com/support/solutions/articles/6000060968-what-outputs-and-sources-does-altmetric-track-, accessed on 16 November, 2020.

In case of Blog platform, Altmetric.com is found to provide higher average blog mentions along with higher coverage as compared to PlumX. Ortega (2018a) has also found that in general Altmetric.com captures more blog mentions per article than PlumX. One may expect that different blog sources tracked by the two aggregators could be the main reason for this. In the overview section, we noted that Altmetric.com tracks about 9,653 blog sources, though information about sources tracked by PlumX is not available. However, the different blog sources tracked could be the main reason for differences in blog mentions captured by Altmetric.com and PlumX, as also indicated by Ortega (2018a).

In terms of correlations between the mentions captured by the two aggregators, the correlation values are higher in case of Twitter and Mendeley platforms, indicating higher agreement between Altmetric.com and PlumX for these platforms. On the other hand, the correlation values are found to be in the lower range in case of Facebook and Blog platform, indicating higher differences in mentions captured by the two aggregators in these platforms. Zahedi & Costas (2018b) have also found highest inter-correlations across the aggregators for Mendeley platform. However, for Twitter platform, the current findings contradict with the finding of Meschede & Siebenlist (2018), where correlation for Twitter platform was found to be in the lower similarity group along with other platforms like Facebook and Blog.

The variations of coverage and mention values of the two aggregators in the four platforms across different document types, subjects and publishers also show interesting patterns. Out of these three aspects, only subject has been explored earlier by Ortega (2020b) for Blog and News mentions. The variations by publisher and document types have not been explored earlier. The present results show that in general Altmetric.com has better coverage and higher mention values than PlumX for Twitter platform across different document types, subjects and publishers. Similarly, PlumX platform is seen to have better coverage and higher mention values than Altmetric.com in case of Facebook, irrespective of the document type, subject and publisher. However, in case of Mendeley and Blog platforms, the coverage and average mention values of the two aggregators do not show a consistent pattern across all the document types, subjects and publishers. In some cases, PlumX has better coverage and higher mentions than Altmetric.com, while in several other cases Altmetric.com has better coverage and higher mentions. Thus, the variations in coverage and mentions across document types, subjects and publishers are more clearly seen in case of Mendeley and Blog platforms.

This study, thus, presents a comprehensive account of variations in coverage and mention values captured by the two aggregators- Altmetric.com and PlumX- across four different platforms. The analysis used a significantly larger dataset as compared to all previous studies. Further, the study is also perhaps the first effort to have analysed the variations across different document types and publishers. The analytical results are interesting and useful and mainly confirm the findings of earlier studies, with some exceptions. The results have practical implications in terms of suggestion for use of a specific aggregator for data from different platforms. For example, based on the results, it can be suggested that one should use Altmetric.com for tracking tweets and Blog mentions, whereas PlumX should be used for Facebook mentions and Mendeley reads. Further, the study also highlights the fact that the two aggregators differ in their sources and data collection and update strategies, which in turn result in varied coverage and altmetric scores captured by them. Therefore, researchers using altmetric data from these aggregators should keep these facts in mind during their analytical studies.

**Conclusion**

The article analysed the coverage and altmetric scores captured by the two altmetric aggregators to identify the similarities and differences between them, and presents following useful results and conclusions. *Firstly*, PlumX has an overall higher coverage of articles than Altmetric.com and tracks a wider number of sources. *Secondly*, Altmetric.com captures more mentions per article in case of Twitter and Blog platform, whereas PlumX captures more mentions per article in case of Mendeley and Facebook platforms. *Thirdly*, correlation values indicate that Altmetric.com and PlumX agree more in their mention values for Twitter and Mendeley platforms but differ more in case of mention values for Facebook and Blog platforms. *Fourthly*, Altmetric.com is found to have better coverage of Twitter, whereas PlumX has better coverage of Facebook, across different document types, subjects and publishers. In case of Mendeley and Blog platforms, variations in patterns across different document types, subjects and publishers do not show a consistent pattern. Overall, the analytical results present a comprehensive account of the variations in coverage and mentions of the two aggregators across four different platforms.

**Acknowledgement**

The authors would like to acknowledge support of Stacy Konkiel, Director of Research Relations at Digital Science for providing access to Altmetric.com and Stephanie Faulkner, Director of Product Management and Operations at Elsevier Research Metrics for providing dashboard access to PlumX data for our research work.

# FIGURES

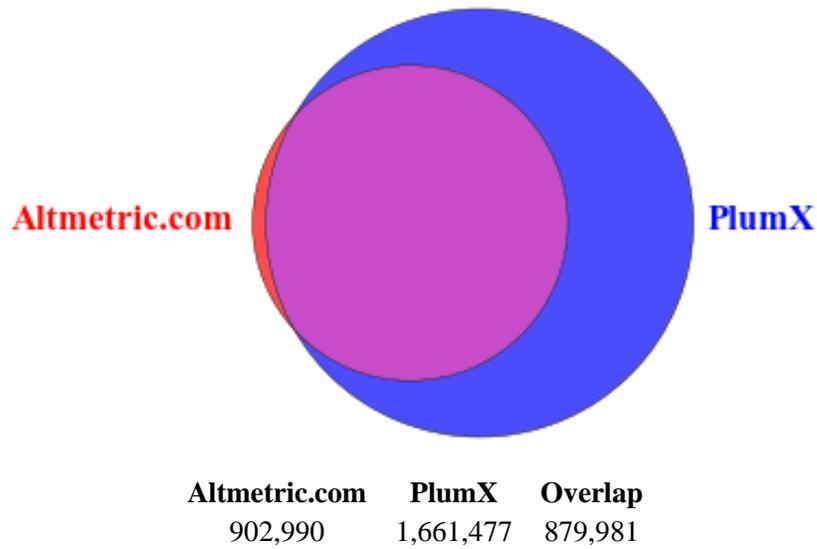

| Altmetric.com | PlumX | Overlap |
|:---:|:---:|:---:|
| 902,990 | 1,661,477 | 879,981 |

**Figure 1: Number of Records found in PlumX and Altmetric.com**

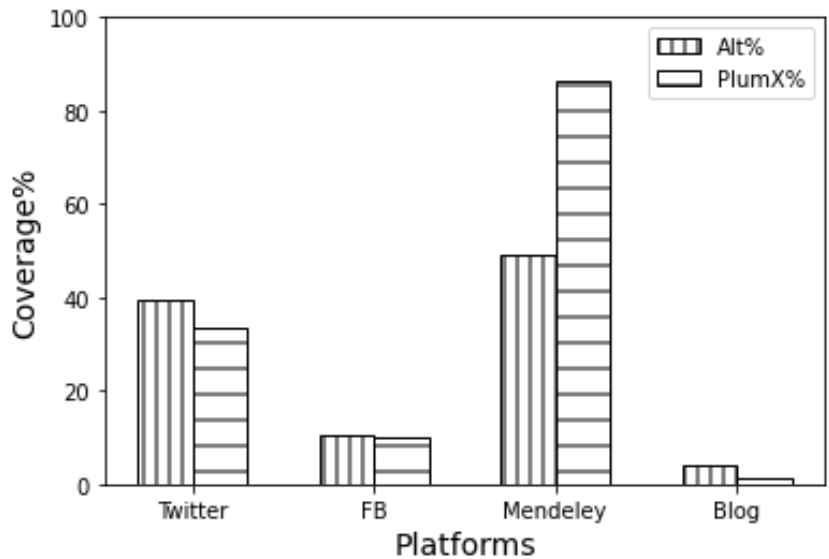

**Figure 2: Coverage Percentages for various platforms in the two aggregators**

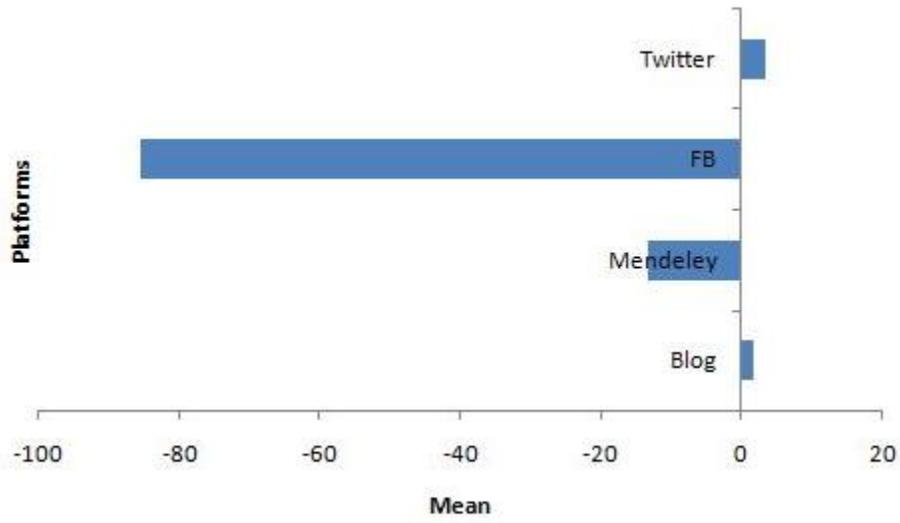

Figure 3: Mean of differences of mentions (Altmetric.com -- PlumX) for different platforms

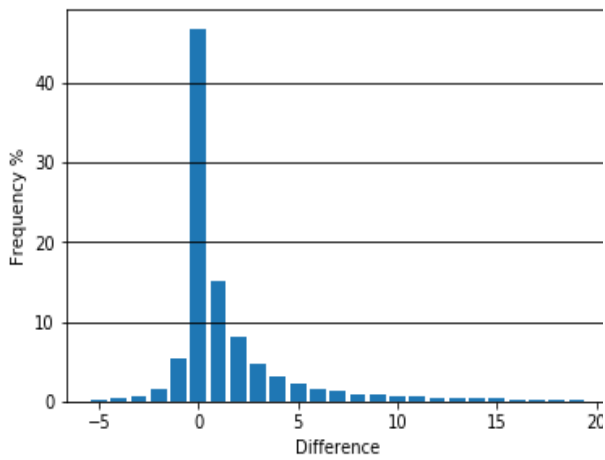

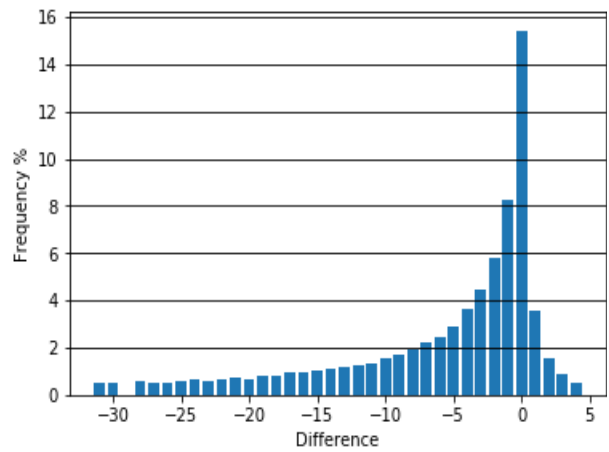

(a) Twitter

(b) Facebook

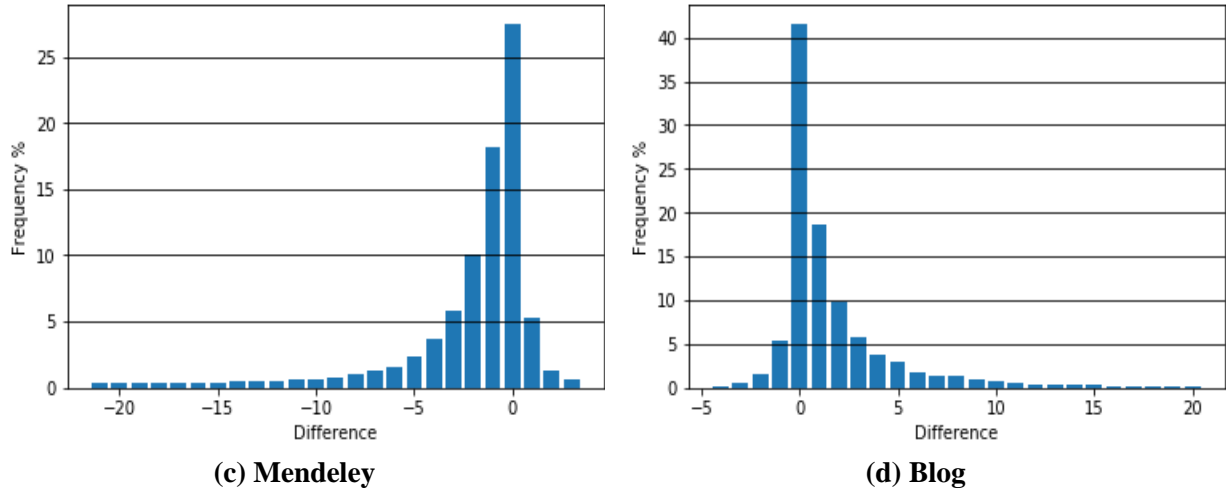

(c) Mendeley  (d) Blog

**Figure 4: Frequency of distribution of differences in mentions (Altmetric.com -- PlumX) for the four platforms.**

# TABLES

**Table 1: Social Media Sources Tracked by the two aggregators**

| Source(s) | Altmetric.com | PlumX |
|---|---|---|
| Twitter | x | x |
| Facebook | x | x |
| YouTube | x | x |
| Reddit | x | x |
| F1000 | x | |
| bit.ly | | x |
| Blog | x | x |
| Figshare | | x |
| GitHub | | x |
| Mendeley | x | x |
| Slideshare | | x |
| SoundCloud | | x |
| SourceForge | | x |
| Vimeo | | x |
| Stack Exchange | | x |
| Stack Overflow | x | |
| Wikipedia | x | x |
| News | x | x |
| Goodreads | | x |
| Amazon | | x |
| Delicious (historical only) | | x |
| CiteULike (historical only) | x | x |
| Dryad | | x |
| DSpace | | x |
| SSRN | | x |
| EBSCO | | x |
| ePrints | | x |
| AiritiiRead eBooks | | x |
| Airiti Library | | x |
| WorldCat | | x |
| LinkedIn | x | |
| Google+ | x | |
| Pinterest | x | |

**Table 2: Bibliographic Impact & Policy Implementation Sources tracked by the two aggregators**

| Source(s) | Altmetric.com | PlumX |
|---|---|---|
| Dynamed Plus Topics | | x |
| Ariti Academic CitationIndex | | x |
| National Institute for Health and Care Excellence (NICE) | | x |
| OJS Journals | | x |
| Open Syllabus | x | |
| Patent Citations | x | |
| Peer Reviews | x | |
| PLOS | | x |
| Policy document source | x | x |
| PubMed Central Europe | | x |
| PubMed Clinical Guidelines | | x |
| PubMedCentral (for PLOS articles only) | | x |
| RePEc | | x |
| SciELO | | x |
| Scopus | | x |
| USPTO | | x |
| Web of Science | x | |
| CrossRef | | x |
| Bepress | | x |
| CABI | | x |
| Dimensions | x | |

**Table 3: Statistics of mentions for different platforms in the two aggregators (only mentions >0)**

| Platfoms | Altmetric.com (902,990) | | | PlumX (1,661,477) | | |
|---|---|---|---|---|---|---|
| | No. of Articles | Mean | Median | No. of Articles | Mean | Median |
| Twitter | 700,985 | 9.56 | 2.0 | 600,051 | 7.46 | 2.0 |
| FB | 185,243 | 2.38 | 1.0 | 182,556 | 47.57 | 5.0 |
| Mendeley | 877,685 | 34.37 | 20.0 | 1,540,214 | 35.08 | 16.0 |
| Blog | 70,387 | 1.92 | 0.0 | 23,052 | 1.64 | 1.0 |

**Table 4: Spearman Rank Correlation between mention counts in the two aggregators**

| PlumX \ Altmetric.com | Twitter | FB | Mendeley | Blog |
|---|---|---|---|---|
| Twitter | 0.823 | 0.330 | 0.353 | 0.261 |
| FB | 0.236 | 0.272 | 0.135 | 0.195 |
| Mendeley | 0.321 | 0.182 | 0.950 | 0.176 |
| Blog | 0.348 | 0.303 | 0.213 | 0.424 |

**Table 5: Document type wise distribution of mentions in the two aggregators**

| Document Type | Aggregator | #Twitter | Avg. Tweets | #FB | Avg. FB Mentions | # Mendeley | Avg Reads | #Blogs | Avg Blog Mentions |
|---|---|---|---|---|---|---|---|---|---|
| Article | Altmetric.com | 545,842 | 8.286 | 140,542 | 2.217 | 691,290 | 33.073 | 55,730 | 1.951 |
| | PlumX | 468,520 | 6.377 | 142,275 | 37.497 | 1,225,475 | 34.218 | 16,910 | 1.646 |
| Book | Altmetric.com | 6,055 | 5.801 | 1,184 | 1.671 | 6,502 | 28.305 | 331 | 1.344 |
| | PlumX | 4,174 | 6.692 | 715 | 23.793 | 12,759 | 23.719 | 148 | 1.405 |
| Book Chapter | Altmetric.com | 2,001 | 5.978 | 307 | 1.651 | 3,259 | 48.541 | 191 | 1.435 |
| | PlumX | 2,043 | 5.38 | 516 | 13.791 | 3,714 | 62.155 | 104 | 1.413 |
| Proceedings Paper | Altmetric.com | 7,089 | 5.784 | 1,796 | 1.744 | 10,832 | 28.63 | 486 | 1.523 |
| | PlumX | 5,763 | 4.61 | 1,785 | 22.769 | 28,649 | 24.647 | 182 | 1.434 |
| Review | Altmetric.com | 58,628 | 12.067 | 17,925 | 2.831 | 67,511 | 68.622 | 5,631 | 1.703 |
| | PlumX | 52,497 | 9.417 | 17,799 | 42.82 | 94,469 | 81.303 | 2,790 | 1.565 |

**Table 6: Discipline wise distribution of mentions in the two aggregators**

| Discipline | Aggregator | #Twitter | Avg. Tweets | #FB | Avg. FB Mentions | # Mendeley | Avg Reads | #Blogs | Avg Blog Mentions |
|---|---|---|---|---|---|---|---|---|---|
| MED | Altmetric.com | 301,704 | 10.505 | 96,605 | 2.587 | 361,663 | 33.583 | 23,116 | 1.946 |
| | PlumX | 266,853 | 7.748 | 78,164 | 49.002 | 516,696 | 40.221 | 9,405 | 1.527 |
| PHY | Altmetric.com | 72,979 | 3.842 | 11,623 | 1.466 | 93,002 | 21.201 | 9,053 | 1.488 |
| | PlumX | 53,269 | 3.292 | 14,049 | 30.972 | 204,802 | 19.522 | 1,143 | 1.85 |
| CHE | Altmetric.com | 55,197 | 3.628 | 9,020 | 1.401 | 72,337 | 31.303 | 3,575 | 1.45 |
| | PlumX | 45,810 | 3.081 | 12,155 | 22.095 | 161,466 | 27.715 | 934 | 1.226 |
| ENV | Altmetric.com | 40,432 | 6.476 | 9,010 | 1.844 | 53,018 | 38.809 | 4,808 | 1.719 |
| | PlumX | 35,354 | 5.585 | 10,501 | 28.799 | 99,051 | 37.798 | 1,476 | 1.732 |
| AH | Altmetric.com | 10,194 | 5.999 | 2,816 | 1.571 | 12,081 | 12.248 | 814 | 1.478 |
| | PlumX | 6,963 | 5.221 | 1,898 | 30.347 | 25,738 | 11.274 | 287 | 1.307 |
| SS | Altmetric.com | 114,537 | 9.84 | 36,918 | 1.961 | 136,109 | 42.023 | 12,726 | 1.84 |
| | PlumX | 95,173 | 8.11 | 28,557 | 28.906 | 192,979 | 49.824 | 4,988 | 1.603 |
| INF | Altmetric.com | 10,854 | 6.278 | 1,658 | 1.486 | 20,271 | 36.987 | 621 | 1.588 |
| | PlumX | 7,423 | 5.735 | 3,202 | 18.685 | 65,892 | 28.539 | 221 | 1.534 |
| BIO | Altmetric.com | 109,393 | 8.239 | 27,229 | 2.056 | 130,175 | 38.03 | 11,260 | 1.783 |
| | PlumX | 102,863 | 6.201 | 26,304 | 33.449 | 181,715 | 41.593 | 3,314 | 1.438 |
| MAR | Altmetric.com | 30,368 | 3.573 | 4,800 | 1.503 | 45,706 | 31.258 | 1,662 | 1.812 |
| | PlumX | 23,781 | 3.009 | 6,854 | 25.608 | 136,295 | 26.78 | 536 | 1.278 |
| MAT | Altmetric.com | 10,772 | 6.138 | 1,227 | 1.516 | 15,289 | 24.91 | 611 | 1.624 |
| | PlumX | 8,433 | 4.769 | 2,251 | 27.961 | 38,415 | 19.445 | 158 | 1.741 |
| GEO | Altmetric.com | 55,972 | 5.952 | 10,348 | 1.884 | 79,222 | 29.949 | 4,713 | 1.64 |
| | PlumX | 48,856 | 4.806 | 12,651 | 26.889 | 143,839 | 31.889 | 1,421 | 1.53 |
| ENG | Altmetric.com | 18,616 | 3.336 | 2,904 | 1.476 | 40,115 | 33.537 | 888 | 1.568 |
| | PlumX | 13,705 | 3.044 | 6,128 | 25.164 | 151,743 | 26.519 | 405 | 1.696 |
| MUL | Altmetric.com | 56,535 | 25.09 | 19,163 | 3.775 | 64,478 | 48.764 | 9,856 | 2.886 |
| | PlumX | 51,136 | 18.882 | 26,362 | 99.871 | 84,051 | 60.863 | 3,634 | 2.117 |
| AGR | Altmetric.com | 29,005 | 4.707 | 7,986 | 1.934 | 38,359 | 33.279 | 2,360 | 1.427 |
| | PlumX | 25,774 | 3.85 | 7,467 | 24.109 | 78,858 | 31.689 | 819 | 1.319 |

Table 7: Publisher wise summary of values of Altmetric.com and PlumX

| Publisher | Aggregator | #Journals | Coverage % | Twitter (%) | Twitter Mean | Facebook (%) | Facebook Mean | Mendeley (%) | Mendeley Mean | Blog (%) | Blog Mean |
|---|---|---|---|---|---|---|---|---|---|---|---|
| Springer | Altmetric.com | 1,512 | 44.47 | 73.5 | 4.98 | 16.4 | 1.66 | 97.5 | 26.24 | 5.1 | 1.45 |
| | PlumX | 1,588 | 93.01 | 28.7 | 3.97 | 10.4 | 28.98 | 94.1 | 24.18 | 0.7 | 1.41 |
| Nature | Altmetric.com | 125 | 79.44 | 86.9 | 31.83 | 29.7 | 4.41 | 98.6 | 64.37 | 18.2 | 2.98 |
| | PlumX | 126 | 97.74 | 63 | 23.55 | 25.2 | 144.87 | 95.5 | 81.84 | 5.1 | 2.22 |
| PLoS | Altmetric.com | 8 | 86.88 | 88 | 12.28 | 22.3 | 2.3 | 98.6 | 36.47 | 10.2 | 2.1 |
| | PlumX | 8 | 98.46 | 72 | 8.96 | 43.1 | 47 | 96.4 | 48.08 | 3 | 1.57 |
| Elsevier | Altmetric.com | 1,836 | 43.10 | 72.3 | 6.67 | 16.5 | 2.01 | 97.7 | 39.85 | 4.9 | 1.67 |
| | PlumX | 1,867 | 94.46 | 31.3 | 5.57 | 12.5 | 27.07 | 94.1 | 40.08 | 0.9 | 1.43 |
| IEEE | Altmetric.com | 163 | 24.98 | 37.7 | 3.31 | 5 | 1.69 | 96.7 | 32.02 | 1.6 | 1.68 |
| | PlumX | 167 | 91.36 | 3.7 | 2.34 | 1 | 17.62 | 97.9 | 24.78 | 0.2 | 4.78 |
| Wiley | Altmetric.com | 1,317 | 64.28 | 79.8 | 7.10 | 24.4 | 1.88 | 97.8 | 31.77 | 6.8 | 1.62 |
| | PlumX | 1,322 | 93.16 | 49.8 | 6.64 | 5 | 31.11 | 93.7 | 36.15 | 1.9 | 1.53 |
| Taylor & Francis | Altmetric.com | 1,353 | 45.83 | 73.9 | 6.29 | 15.8 | 1.88 | 96.2 | 25.22 | 4.3 | 1.51 |
| | PlumX | 1,374 | 91.42 | 31.5 | 5.3 | 3.2 | 14.35 | 92.9 | 23.88 | 0.9 | 1.29 |
| ACM | Altmetric.com | 31 | 54.78 | 75.8 | 5.12 | 19.3 | 1.25 | 94.7 | 23.85 | 8.7 | 1.29 |
| | PlumX | 32 | 94.52 | 23 | 4.35 | 6.7 | 14.59 | 92.2 | 23.22 | 0.7 | 1 |
| IOP | Altmetric.com | 69 | 54.13 | 79 | 6.13 | 13 | 1.67 | 97.3 | 18.03 | 24 | 1.63 |
| | PlumX | 71 | 90.22 | 29.1 | 6.8 | 8.1 | 28.69 | 92.2 | 19.16 | 1.1 | 1.86 |
| Oxford Univ Press | Altmetric.com | 325 | 68.66 | 86.4 | 10.29 | 21.9 | 2.03 | 96.3 | 33.87 | 14.1 | 1.54 |
| | PlumX | 327 | 92.17 | 45 | 5.73 | 10.5 | 23.49 | 88.7 | 41.15 | 1.7 | 1.38 |
| SAGE | Altmetric.com | 660 | 61.05 | 81.1 | 8.81 | 20.3 | 2.02 | 97.2 | 32.70 | 8.7 | 1.81 |
| | PlumX | 663 | 94.73 | 41.2 | 8.31 | 3.5 | 25.91 | 92.4 | 36.3 | 2.3 | 1.58 |
| Hindawi | Altmetric.com | 87 | 35.21 | 68.6 | 3.1 | 13.3 | 2.27 | 98.3 | 27.22 | 3.1 | 1.15 |
| | PlumX | 97 | 93.17 | 26.9 | 2.76 | 13.8 | 20.09 | 95.7 | 23.85 | 0.6 | 1.19 |
| Cell | Altmetric.com | 38 | 70.01 | 90.7 | 22.23 | 32.5 | 2.93 | 96.2 | 91.59 | 20.5 | 2.65 |
| | PlumX | 38 | 94.9 | 64.6 | 14.84 | 24.6 | 54.76 | 80.1 | 120.58 | 4.5 | 1.54 |
| MDPI | Altmetric.com | 55 | 69.34 | 74.3 | 5.45 | 10.3 | 2.49 | 99.3 | 32.30 | 3.2 | 1.44 |
| | PlumX | 55 | 97.35 | 44.1 | 4.84 | 22.2 | 32.9 | 98 | 34.51 | 1.5 | 1.32 |
| Cambridge Univ Press | Altmetric.com | 305 | 46.50 | 75.1 | 7.01 | 22.8 | 1.70 | 91.4 | 25.61 | 8.3 | 1.55 |
| | PlumX | 311 | 84.1 | 25.8 | 4.65 | 7.5 | 40.68 | 83 | 26.71 | 0.7 | 1.29 |
| Emerald | Altmetric.com | 89 | 30.49 | 48.3 | 3.42 | 8.1 | 1.61 | 97.8 | 51.13 | 1.7 | 1.31 |
| | PlumX | 91 | 92.52 | 15.5 | 2.94 | 0.6 | 1 | 98.5 | 58.48 | 0.4 | 4.37 |